\documentclass[usenatbib]{mn2e}

\usepackage{times}
\usepackage{lscape}
\usepackage{subfigure}
\usepackage[normalem]{ulem}
\usepackage{xcolor}
\input{psfig.sty}

\def\gsim{\mathrel{\raise0.35ex\hbox{$\scriptstyle >$}\kern-0.6em\lower0.40ex\hbox{{$\scriptstyle \sim$}}}}
\def\lsim{\mathrel{\raise0.35ex\hbox{$\scriptstyle <$}\kern-0.6em\lower0.40ex\hbox{{$\scriptstyle \sim$}}}}
\def\apj{ApJ}
\def\apjl{ApJL}
\def\mnras{MNRAS}

\def\apss{APSS}

\def\aapr{A\&Arev}
\def\aap{A\&A}
\def\aj{AJ}
\def\apjs{ApJS}

\def\nat{Nature}

\makeatother

\title[MUSE observations of TN\,J1338$-$1942 at {\emph
    z}\,=\,4.1]{Mapping the Dynamics of a Giant Ly$\alpha$ Halo at
  {\emph z}\,=\,4.1 with MUSE: The Energetics of a Large Scale
  AGN-Driven Outflow around a Massive, High-Redshift Galaxy}

\author[Swinbank et al.]
{ \parbox[h]{\textwidth}{
A.\,M.\ Swinbank,$^{\! 1,*}$
J.\,D.\,R.\ Vernet,$^{\! 2}$
Ian Smail,$^{\! 1}$
C.\ De Breuck,$^{\! 2}$
R.\ Bacon$^{3}$,
T. Contini$^{4}$,
J. Richard$^{3}$,
H.\,J.\,A.\ R\"ottgering$^{5}$,
T. Urrutia$^{6}$,
\& B.\ Venemans$^{7}$
}
\vspace*{6pt}\\
$^{1}$Institute for Computational Cosmology, Durham University, South Road, Durham DH1 3LE UK\\
$^{2}$European Southern Observatory, Karl Schwarzschild Stra{\ss}e 2, 85748, Garching, Germany \\
$^{3}$CRAL, Observatoire de Lyon, Universite Lyon 1, 9 Avenue Ch. Andre, F-69561 Saint Genis Laval Cedex, France\\
$^{4}$Dipartimento di Fisicae Astronomia, University of Padova, Vicolo dell'Osservatorio 2, 35133, Padova, Italy\\
$^{5}$Leiden Observatory, Leiden University, PO Box 9513, 2300 RA Leiden, The Netherlands\\
$^{6}$Leibniz Institut f\"ur Astrophysik, An der Sternwarte 16, 14482 Potsdam, Germany \\
$^{7}$Max-Planck Institut f\"ur Astronomie, K\"onigstuhl 17, 69117, Heidelberg, Germany\\
$^{*}$ email: a.m.swinbank@durham.ac.uk \\
}

\begin{document}

\maketitle

\begin{abstract}
We present deep MUSE integral-field unit (IFU) spectroscopic
observations of the $\sim$\,150\,kpc Ly$\alpha$ halo around the
$z$\,=\,4.1 radio galaxy TNJ\,J1338$-$1942.  This 9-hr observation
maps the two-dimensional kinematics of the Ly$\alpha$ emission across
the halo.  We identify two H{\sc i} absorbers which are seen against
the Ly$\alpha$ emission, both of which cover the full
$\sim$\,150\,$\times$\,80\,kpc extent of the halo and so have covering
fractions $\sim$\,1.  The stronger and more blue-shifted absorber
($\Delta v\sim $\,$-$1200\,km\,s$^{-1}$ from the systemic) has
dynamics that mirror that of the underlying halo emission and we
suggest that this high column material ($n$(H{\sc
  i})\,$\sim$\,10$^{19.4}$\,cm$^{-2}$), which is also seen in C{\sc
  iv} absorption, represents an out-flowing shell that has been driven
by the AGN or the star formation within the galaxy.  The weaker
($n$(H{\sc i})\,$\sim$\,10$^{14}$\,cm$^{-2}$) and less blue shifted
($\Delta v\sim $\,$-$500\,km\,s$^{-1}$) absorber most likely
represents material in the cavity between the out-flowing shell and
the Ly$\alpha$ halo.  We estimate that the mass in the shell must be
of order 10$^{10}$\,M$_{\odot}$ -- a significant fraction of the ISM
from a galaxy at $z$\,=\,4.  The large scales of these coherent
structures illustrate the potentially powerful influence of AGN
feedback on the distribution and energetics of material in their
surroundings.  Indeed, the discovery of high-velocity
($\sim$\,1000\,km\,s$^{-1}$), group-halo-scale (i.e. $>$\,150kpc) and
mass-loaded winds in the vicinity of the central radio source are
broadly in agreement with the requirements of models that invoke
AGN-driven outflows to regulate star formation and black-hole growth
in massive galaxies.
\end{abstract}


\section{Introduction}

Theoretical models of galaxy formation are increasingly successful in
their ability to predict the observed properties of galaxies across
cosmic time.  The most successful models invoke a dramatic form of
energy injection from the central active galactic nuclei (AGN) and
from star-formation and supernovae to drive high-velocity (100's
km\,s$^{-1}$), galaxy-wide (i.e. $\gg$\,10kpc) and mass-loaded
energetic outflows.  These winds expel significant amounts of gas from
the host galaxy and consequently this shuts down future star formation
and black hole growth.  This process should also enrich the larger
scale environment with metals
\citep[e.g.\ ][]{Silk98,Fabian99,Benson03,Granato04,DiMatteo05,Springel03,Springel05b,Hopkins06b,Hopkins08,Booth10,Debuhr12}.
In the models, this feedback mechanism is required to prevent the
run-away cooling of gas in massive halos and so match the galaxy
luminosity function at $z$\,=\,0 \citep[e.g.\ ][]{Bower06,Croton06}.

If this feedback mechanism is the dominant route by which massive
galaxies terminate their star formation, then it is critical that the
AGN feedback is at its maximum efficiency at the epoch where the bulk
of the stars in the most luminous elliptical galaxies appear to be
formed, $z>$\,3--5 \citep{Nelan05}.  This is therefore the era to
search for evidence of AGN feedback in massive galaxies.
High-redshift radio galaxies are one potential laboratory to search
for evidence of this feedback ``in action'', as they appear to be
amongst the most evolved and most massive galaxies in the early
Universe, host energetic AGN, and have the properties expected for the
progenitors of today's central cluster galaxies \citep{deBreuck02}.

Potential evidence for such ``AGN feedback'' may come from the
discovery of very extended Ly$\alpha$ halos with spatial extents of
$\sim$\,100--200\,kpc which are being found in increasing numbers in
the high-redshift Universe, including around high-redshift radio
galaxies \citep{vanojik97,Venemans07}.  Many of these appear to be
associated with young, massive galaxies which display high-levels of
starburst and AGN activity \citep[e.g.\ ][]{vanBreugel98,Matsuda11}.
One simple interpretation is that these Ly$\alpha$ halos represent gas
reservoirs which are being ionised by UV emission, shocks and outflows
from the active galaxies within the halos
\citep[e.g.\ ][]{Taniguchi01,Geach09b,Smail13}.  However, theoretical models
have also suggested that these halos may instead be powered by cooling
from gas in so-called ``cold-flows''
\citep[e.g.\ ][]{Fardal01,Dekel09}.  The spatially resolved dynamics
and line profiles of these Ly$\alpha$ halos, as well as their chemical
enrichment should be able to differentiate which
mechanism is responsible for their formation \citep{Bower04,Smail13}.

The two-dimensional Ly$\alpha$ surface brightness distributions
\citep[e.g.\ ][]{Matsuda12} or one-dimensional kinematics of
high-redshift radio galaxies
\citep[e.g.\ ][]{vanojik97,Binette00,VillarMartin03,Ohyama04,Matsuda06,Humphrey07,Humphrey09,Yang11,Yang14}
suggest that the central AGN plays an important role in shaping the
environments around massive, high redshift galaxies on $\gsim$\,10's
of kilo-parsec scales.  For example, observations have reveals that
the spatial extent of the Ly$\alpha$-emitting gas is correlated with
the spatial scales of the jets, whilst the distortion seen in the
Ly$\alpha$ morphology is also typically correlated with the distortion
in the radio emission (the extended emission line region morphology
always appears to be brightest on the side of the radio lobes that is
closer to the nucleus; \citealt{McCarthy91,Tadhunter00}), suggesting
that powerful radio sources are (at least partly) responsible for
environmental effects on large scales.  In terms of the large scale
environment, \citet{vanojik97} showed that the galaxies with radio
jets that are extended on scales smaller than $\lsim$\,50\,kpc have
the strongest strong H{\sc i} absorption seen against Ly$\alpha$ (and
likewise, those with radio extents $\gsim$\,50\,kpc do not show strong
absorbers).  One explanation is that when radio sources growth they
destroy the H{\sc i}.  However, this model also requires a duty cycle
in which the ``feedback'' expels gas, which cools then to form a shell
of H{\sc i}.  A subsequent ``feedback'' event then ionises the gas.
This model may also explain why the dynamics of the gas in the inner
halo (located within the radio source) tend to be turbulent
($\sigma\sim$\,700--1500\,km\,s$^{-1}$) whilst the out halo is more
quiescent ($\sigma\sim$\,300\,km\,s$^{-1}$; \citealt{vanojik96}).

To measure the energetics of the Ly$\alpha$ emission, search for and
place dynamical constraints on the out-flowing material requires
resolved spectroscopy.  However, relatively few galaxies have been
studied using their two-dimensional Ly$\alpha$ velocity field, as
required to truly understand their origin.  The first attempt to
derive the morphologies and dynamics of extended Ly$\alpha$ halos
were made by \citet{Hippelein93} who found a coherent and spatially
extended ($\sim$100\,kpc$^2$) and moderate column, $n$(H{\sc
  i})$\sim$10$^{15}$\,cm$^{2}$, H{\sc i} absorber in front of a 4C
radio galaxy at $z$\,=\,3.8 \citep[see also][]{Adam97}.  However,
relatively few galaxies have been studied using their two-dimensional
Ly$\alpha$ velocity field, as required to truly understand their
origin \citep[e.g.\ see][for a few
  examples]{Bower04,Wilman05,VillarMartin07a,Sanchez09,Weijmans10,Martin14}.
In part, this is due to the high redshift and hence large luminosity
distances required to redshift Ly$\alpha$ into the optical band, which
means that spatially resolving the low surface brightness features in
the outer halo have, until now, been beyond the sensitivity limits of
most instruments (even on 8-meter telescopes).  However, with the
commissioning of the sensitive Multi-Unit Spectroscopic Explorer
(MUSE) on the ESO\,/\,VLT, it is now possible to carry out these
observations and so determine the nature of extended Ly$\alpha$ halos.

In this paper, we exploit MUSE observations to determine the nature of
extended Ly$\alpha$ halo around a powerful radio galaxy,
TNJ\,J1338$-$1942 at $z$\,=\,4.1 (hereafter TNJ\,J1338).  TNJ\,1338 was
identified as hosting a giant Ly$\alpha$ halo by \citet{Venemans02}.
In a subsequent paper, \citet{Venemans07} \citep[see also][]{Miley04}
identified twenty Ly$\alpha$ emitters within a projected distance of
1.3\,Mpc and 600\,km\,s$^{-1}$ of the radio galaxy (an over-density of
4--15 compared to the average field at this redshift), suggesting a
mass for the structure of M$\sim$5\,$\times$\,10$^{14}$M$_{\odot}$
which makes structure a likely ancestor of a massive galaxy cluster at
the present day (with the radio galaxy likely to be a brightest
cluster galaxy in formation).  This high-redshift radio galaxy remains
one of the the brightest and most luminous known in Ly$\alpha$
($L_{\rm Ly\alpha}\sim $\,5\,$\times$\,10$^{44}$\,erg\,s$^{-1}$).
Both its Ly$\alpha$ profile and radio structure are very asymmetric
\citep{DeBreuck99,Wilman04}, which indicates strong interaction with
dense gas, and the rest-frame radio luminosity is comparable to that
of the most luminous 3C radio sources (P$_{\rm
  178\,MHz}\sim$\,4\,$\times$\,10$^{35}$\,erg\,s$^{-1}$\,Hz$^{-1}$\,sr$^{-1}$).
The MUSE observations presented here allow us to map the emission
within the $\sim$\,150-kpc long Ly$\alpha$ halo, matching the
sensitivity of previous narrow-band Ly$\alpha$ imaging of this
structure, but with the {\it crucial} advantage of high spectral
resolution kinematics, necessary to differentiate the signatures of
infall or outflow \citep{Adams09,Bower04,Wilman05}.

The structure of this paper is as follows: in \S~2 we present the MUSE
observations and data-reduction.  In \S~3 we discuss the morphology
and energetics of the Ly$\alpha$ halo and those of the H{\sc i}
absorbers seen against the underlying emission.  In \S4 we present our
conclusions.  Throughout this paper, we adopt a cosmology with
$\Omega_{\rm m}$\,=\,0.27, $\Omega_\Lambda$\,=\,0.73 and
$H_0$\,=\,71\,km\,s$^{-1}$\,Mpc$^{-1}$ giving an angular scale of
7.0\,kpc\,arcsec$^{-1}$ at $z$\,=\,4.1.  

%
%
\begin{figure*}
  \centerline{\psfig{file=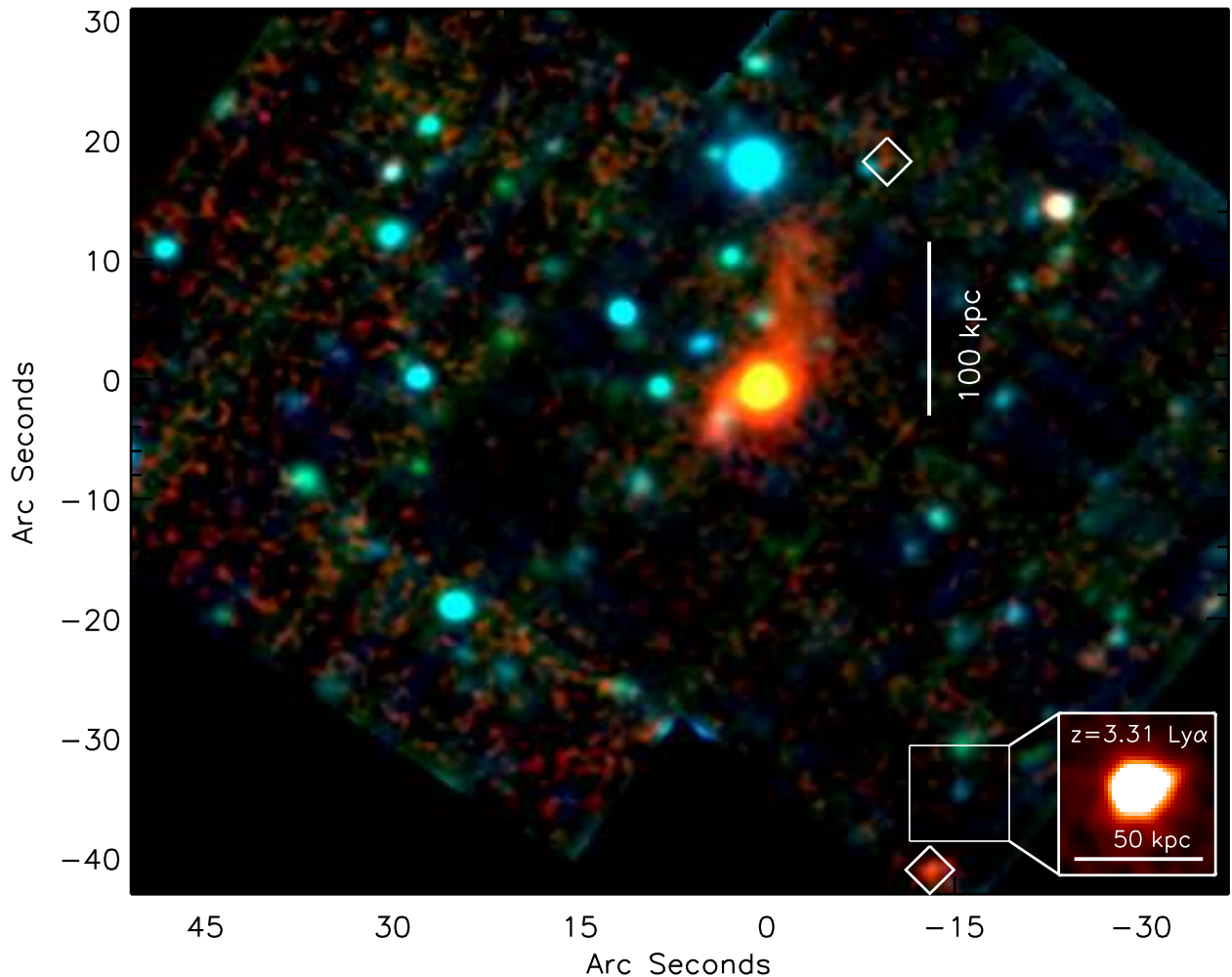,width=7in,angle=90}}
\caption{Colour, ($V$, $I$, Ly$\alpha_{z=4.1}$) image of the field
  around TNJ\,J1338 from the MUSE observations.  This colour image
  illustrates the full extent the giant halo around the $z=$\,4.1
  radio galaxy.  Moreover, a search for other emission lines within
  the MUSE cube identifies a new giant ($\sim$\,60\,kpc) halo around a
  foreground $z$\,=\,3.31 galaxy located to the south west of the
  radio source. We also mark the positions of the two $z$\,=\,4.1
  galaxies in the halo of the radio galaxy (open diamonds).  The
  origin of the image is centered on the radio galaxy at
  $\alpha$:\,13:38:26.1, $\delta$:\,$-$19:42:30.5 (J2000).}
\label{fig:colimg}
\end{figure*}

%
%
\begin{figure}
  \centerline{\psfig{file=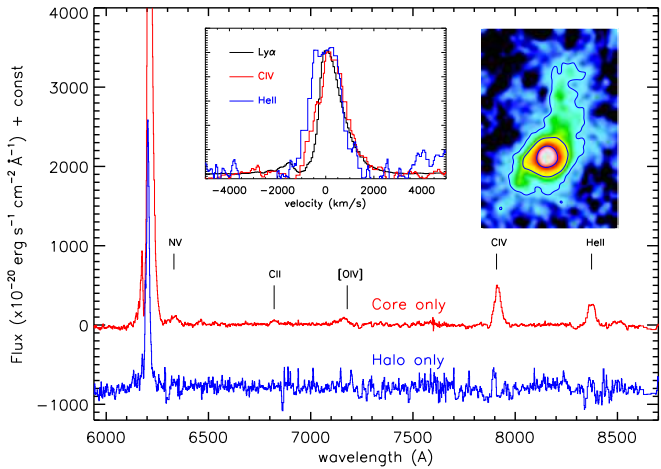,width=3.6in,angle=90}}
\caption{The spectrum extracted from the core of the Ly$\alpha$ halo
  compared to that of the extended halo.  The core spectrum shows
  strong emission lines associated with the AGN activity (e.g.\ N{\sc
    v}, [O{\sc iv}], C{\sc iv} and He{\sc ii}).  The emission from the
  halo (which should exclude any seeing-related contamination from the
  core, but will not exclude scattered emission within the halo) also
  shows weak C{\sc iv} emission.  The Ly$\alpha$ image of the halo
  ({\it inset}) shows the two regions from which the core and halo
  were extracted (for the core, all pixels within the inner contour
  were included and for the halo, all pixels between the second and
  third contours were included).  The second inset compares the line
  profile of the Ly$\alpha$, C{\sc iv} and He{\sc ii} from the core
  region.  In this panel, we use the He{\sc ii} emission to define the
  systemic redshift, $z$\,=\,4.1057 (and transform the Ly$\alpha$,
  C{\sc iv} and He{\sc ii} to the same rest-frame velocity scale).  In
  the core region, the Ly$\alpha$ and C{\sc iv} share the same
  characteristic (asymmetric) line profile, implying the presence of
  C{\sc iv} in the absorbing material and suggesting that any
  absorbing materials must be enriched in metals}
\label{fig:CoreHaloSpec}
\end{figure}

%
%
\begin{figure*}
  \centerline{\psfig{file=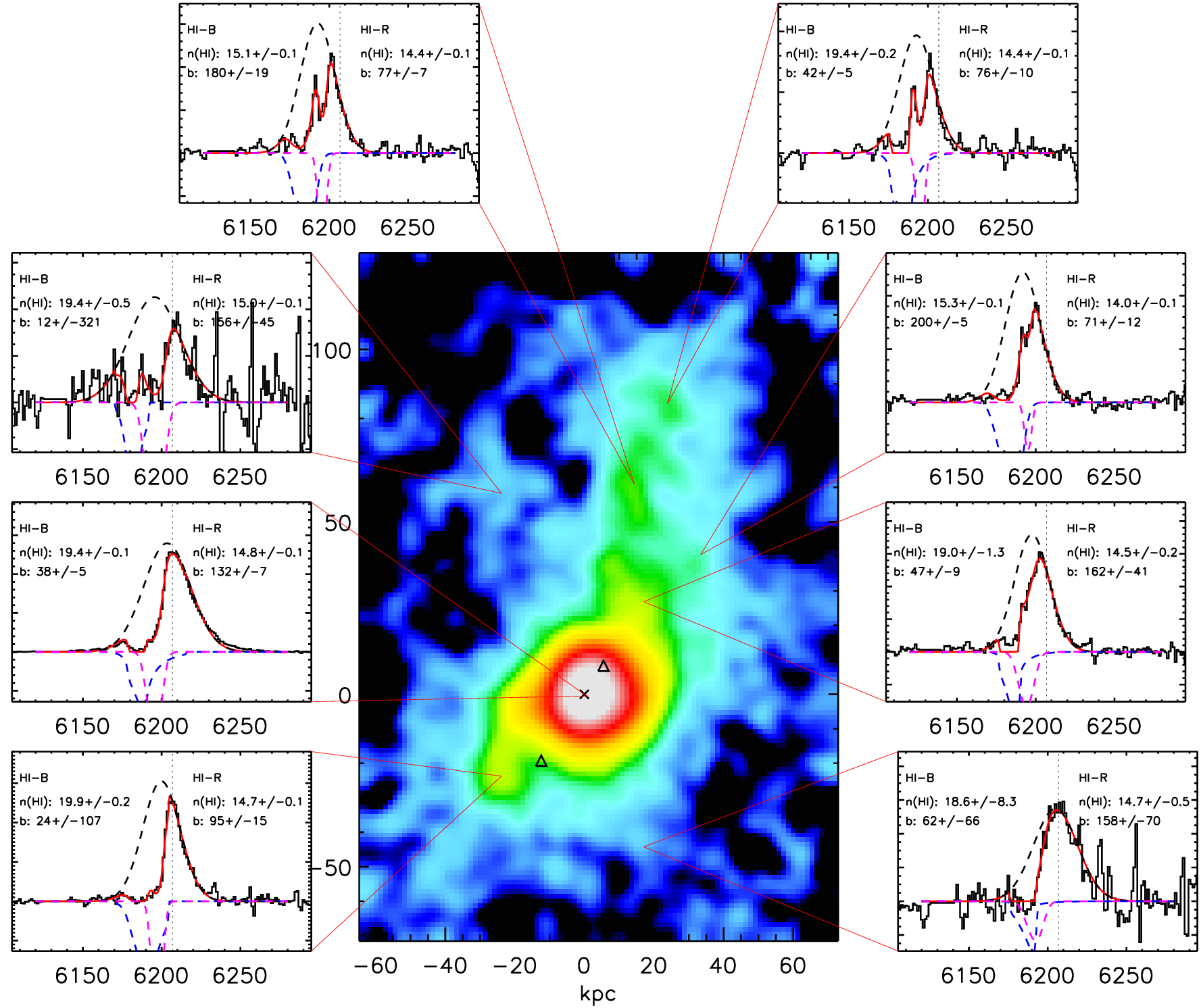,width=7in,angle=0}}
  \caption{{\it Upper panel:} Continuum subtracted, wavelength
    collapsed, narrow-band image of the TNJ\,J1338 Ly$\alpha$ halo from
    the MUSE cube (collapsed over the wavelength range 6100--6300\AA).
    The cross denote the position of the UV-continuum (and radio)
    counterparts, and the open triangles show the positions of the
    radio hot-spots (end points of the jets) from \citet{DeBreuck99}
    indicating the compact size of the radio source compared to the
    highly-extended Ly$\alpha$ halo.  We also show example spectra of
    regions within the halo.  These are typically extracted from
    2$''\times$\,2$''$ regions, and in all cases, the line profile is
    best-fit by two H{\sc-i} absorbers superimposed on an underlying
    Gaussian emission profile.  Both absorbers have a covering factor
    of unity; the first absorber, H{\sc i}-B appears blue-shifted from
    the (underlying) Gaussian emission line profile by
    $\sim$\,$-$700\,km\,s$^{-1}$ and has a column density of $n$(H{\sc
      i})$\sim$\,10$^{19.3}$\,cm$^{-2}$ whilst the second absorber,
    H{\sc i}-R is much closer to the underlying halo emission, $\Delta
    v\sim$\,$-$200\,km\,s$^{-1}$ but has a much lower column density,
    $n$(H{\sc i})$\sim$\,10$^{14.5}$\,cm$^{-2}$.  }
\label{fig:ImageSpec}
\end{figure*}

\section{Observations and Data Reduction}

We observed TNJ\,J1338 with MUSE for a total of 9.0\,hours between 2014
April 30 and 2014 June 30 during a combination of commission and
Science Verification time.  The MUSE IFU provides full spectral
coverage spanning 4770--9300\AA, a contiguous field of view of
60\,$''\times$\,60\,$''$, with a spatial sampling of
0.2$''$\,pixel$^{-1}$ and a spectral resolution of $R$\,=\,3000 at
$\lambda$\,=\,6200\AA\, (the wavelength of the Ly$\alpha$ at
$z$\,=\,4.1).  The observations were carried out in dark time, good
sky transparency and 0.7$''$ $V$-band seeing.  Each 1\,hr observation
was split in to 1800\,second exposures with a small (2$''$) dither.
We obtained two overlapping pointings; the first centered at
$\alpha$:\,13\,38\,28.0 $\delta$:\,$-$19\,42\,24 and a second, offset
by approximately 30$''$ at $\alpha$:\,13\,38\,25.0
$\delta$:\,$-$19\,42\,32.  Both pointings cover the $z$\,=\,4.1 radio
galaxy and the final field of view is 1.5$'\times$\,1.0$'$
(Fig.~\ref{fig:colimg}).

To reduce the data, we use the MUSE pipeline through {\sc esorex}
which wavelength calibrates, flat-fields the spectra and forms the
datacube \citep{Weilbacher14}.  Due to the excellent flat-field
stability of MUSE, sky subtraction was performed on each individual
cube by identifying and subtracting the sky emission using blank areas
of sky at each wavelength slice (i.e. no additional sky-only
observations are required), and the final mosaic was then constructed
using an average with a 3-$\sigma$ clip to reject cosmic rays after
registering the stars in continuum images from each individual cube.

A visual inspection of the final datacube reveals a very strong
detection of the extended Ly$\alpha$ emission from the $z$\,=\,4.1
halo at 6200\AA\, as well as Ly$\alpha$ emission from a number of
galaxies at the same redshift.  In Fig.~\ref{fig:colimg} we show a
colour image of the field generated from the MUSE cube.  In this
image, we use the continuum from the wavelength ranges 4900--6900\AA\,
and 6900--9000\AA\, as the blue and green images respectively, whilst
the red image is generated by collapsing the cube over the wavelength
range 6100--6200\AA\ (which covers the Ly$\alpha$ from the $z$\,=\,4.1
radio galaxy).  The strong, extended Ly$\alpha$ emission can clearly
be seen, with a spatial extent of $\sim$\,150\,$\times$\,80\,kpc.

We note that a search for line emitters within the cube also detects a
second strong, spatially-extended Ly$\alpha$ halo around a red
$z$\,=\,3.31 galaxy located approximately $\sim$\,30$''$ to the south
west of TNJ\,J1338 (Fig.~\ref{fig:colimg}) which appears to lie in an
over-density comprising a number of other star-forming galaxies and
AGN at this redshift.  In addition, a large number of foreground
[O{\sc ii}], [O{\sc iii}] and Ly$\alpha$ emitters are detected in this
volume.  These will be discussed in a subsequent paper (Urrutia et
al.\ in prep).

\section{Analysis \& Discussion}
\label{sec:analysis}

\subsection{Morphology and Energetics of the Ly$\alpha$ Halo}
\label{sec:morphology}

The morphology of the Ly$\alpha$ emission (Fig.~\ref{fig:colimg}) is
very similar to that seen in the FORS\,2 narrow-band imaging from
\citet{Venemans07}, which has similar surface brightness limits as the
Ly$\alpha$ narrow-band image created from the MUSE cube (collapsing
the cube over 120\AA\ around the Ly$\alpha$ yields an image with a
5-$\sigma$ line flux detection in a 1.2$''$ aperture of $f_{\rm
  5\sigma}$\,=\,$\sim$8\,$\times$\,10$^{-18}$\,erg\,s$^{-1}$\,cm$^{-2}$
which is similar to that of the narrow-band image from
\citealt{Venemans07}; $f_{\rm
  5\sigma}$\,=\,6\,$\times$\,10$^{-18}$\,erg\,s$^{-1}$\,cm$^{-2}$).
Indeed, the morphology of the source shows a strong, spatially
extended core which has a diameter of approximately 5$''$ with a lower
surface brightness halo which is extended along a roughly North--South
direction which extends over a $\sim$\,20$''$ (150\,kpc) in extent.
The elongated Ly$\alpha$ morphology is also consistent with the
direction of the radio jets (Fig.~\ref{fig:colimg}), although the
strongest jet emission is contained within the central
$\sim$\,10\,kpc, which may be indicative of previous AGN activity
(which ionised the extended halo and which is now seen in Ly$\alpha$
emission as the gas cools).  We note that the elongated morphology of
the Ly$\alpha$ -- in the same direction as the radio emission -- is
consistent with that seen in a number of other high-redshift radio
galaxies \citep{Rottgering99conf}.

As an order-of-magnitude estimate of the total H{\sc ii} mass
responsible for the Ly$\alpha$ emitting halo, we can assume case B
recombination and that the gas is at a temperature of 10$^4$\,K, and
following \citet{DeBreuck99}, using the total Ly$\alpha$ emission
intensity (corrected for absorption, see \S~\ref{sec:HI-Dynamics}),
the total H{\sc ii} mass is approximately M(H{\sc
  ii})\,$\sim$\,10$^9$($f_5$\,$L_{44}$\,$V_{70}$)$^{0.5}$\,M$_\odot$,
where $f_5$ is the volume filling factor of 10$^{-5}$
\citep{McCarthy90}, $L_{44}$ is the Ly$\alpha$ luminosity in units of
10$^{44}$\,erg\,s$^{-1}$ and $V_{70}$ is the volume of the region in
units of 10$^{70}$\,cm$^3$.  Adopting a size of
150\,$\times$\,80\,$\times$\,80\,kpc we derive M(H{\sc
  ii})\,$\sim$\,4\,$\times$\,10$^9$\,M$_{\odot}$.

The spatial extent of the Ly$\alpha$ halo in TNJ\,J1338 is comparable
to that seen in a number of other high-redshift radio galaxies
\citep{vanojik97,Humphrey07}.  Indeed, extended Ly$\alpha$ halos are
common around high-redshift radio galaxies, extending up to
$\sim$\,250\,kpc -- far beyond the continuum structures revealed by
{\it HST} images \citep[e.g.\ ][]{VillarMartin03}.  Moreover, the
spatial extent of the Ly$\alpha$ emitting halo around high-redshift
radio galaxies also appear to correlate with the size of the radio
emission \citep{vanojik97} (although of course, this is likely to be a
selection effect since the larger halos are likely to be more easily
detectable around powerful active galaxies since the active nucleus
provides a more intense supply of ionising photons).  Blind
narrow-band surveys have also identified giant Ly$\alpha$ emitting
halos around star-forming galaxies with (relatively) low luminosity
AGNs \citep{Matsuda11} as well as in structures containing multiple
radio-quiet quasars \citep[e.g.\ ][]{Cantalupo14}, suggesting that
while an AGN may play an important role in providing the ionising
source of radiation to ionise the halo, it is not a requirement.
Indeed, the Ly$\alpha$ blobs (LABs) also have spatial extents up to
180\,kpc and the Ly$\alpha$ emission can be attributed to either
photoionisation from the AGN or star-formation from the host galaxies
\citep{Geach09b}.

What powers the Ly$\alpha$ emission in the extended halo of TNJ\,1338
is unclear.  As Fig.~\ref{fig:colimg} shows, the Ly$\alpha$ halo
extends over $\sim$\,10$\times$ that of the radio jets detectable at
GHz radio frequencies, which are contained within the central
$\sim$\,10\,kpc.  One possible interpretation is that the emission in
the extended halo is powered by the cooling of pristine gas within a
dark matter halo \citep{Fardal01,Nilsson06}.  However, the
far-infrared luminosity of TNJ\,1338 implied from {\it
  Herschel}\,/\,SPIRE and JCMT\,/\,SCUBA-850$\mu$m observations is
$L_{\rm IR}$\,=\,7\,$\pm$\,1\,$\times$\,10$^{12}$\,L$_{\odot}$
(SFR\,=\,700\,M$_{\odot}$\,yr$^{-1}$; \citealt{Drouart14}) which gives
a ratio of total Ly$\alpha$\,/\,IR luminosity of
$\sim$1\,$\times$\,10$^{45}$\,/\,4\,$\times$\,10$^{46}$\,=\,0.03.
This luminosity ratio is similar to that from the AGN luminosity
(L$_{\rm AGN}\sim$\,2\,$\times$\,10$^{46}$\,erg\,s$^{-1}$;
\citealt{Drouart14}) which gives L$_{\rm Ly\alpha}$\,/\,L$_{\rm
  AGN}$\,=\,0.05.  Thus, only a small fraction of either the infrared
(star-formation) or AGN luminosity is required to power the Ly$\alpha$
emission.  However, we note that the high Ly$\alpha$\,/\,He{\sc ii}
ratio averaged over the halo is Ly$\alpha$\,/\,He{\sc
  ii}$\lambda$1640\,=\,18.7\,$\pm$\,0.5 (see \S~\ref{sec:resolved})
which is difficult to explain with AGN photo-ionisation only
\citep{VillarMartin07b}, and suggests we are likely to be observing a
combination of energy sources that are heating the halo.

%
%
\begin{figure*}
  \centerline{\psfig{file=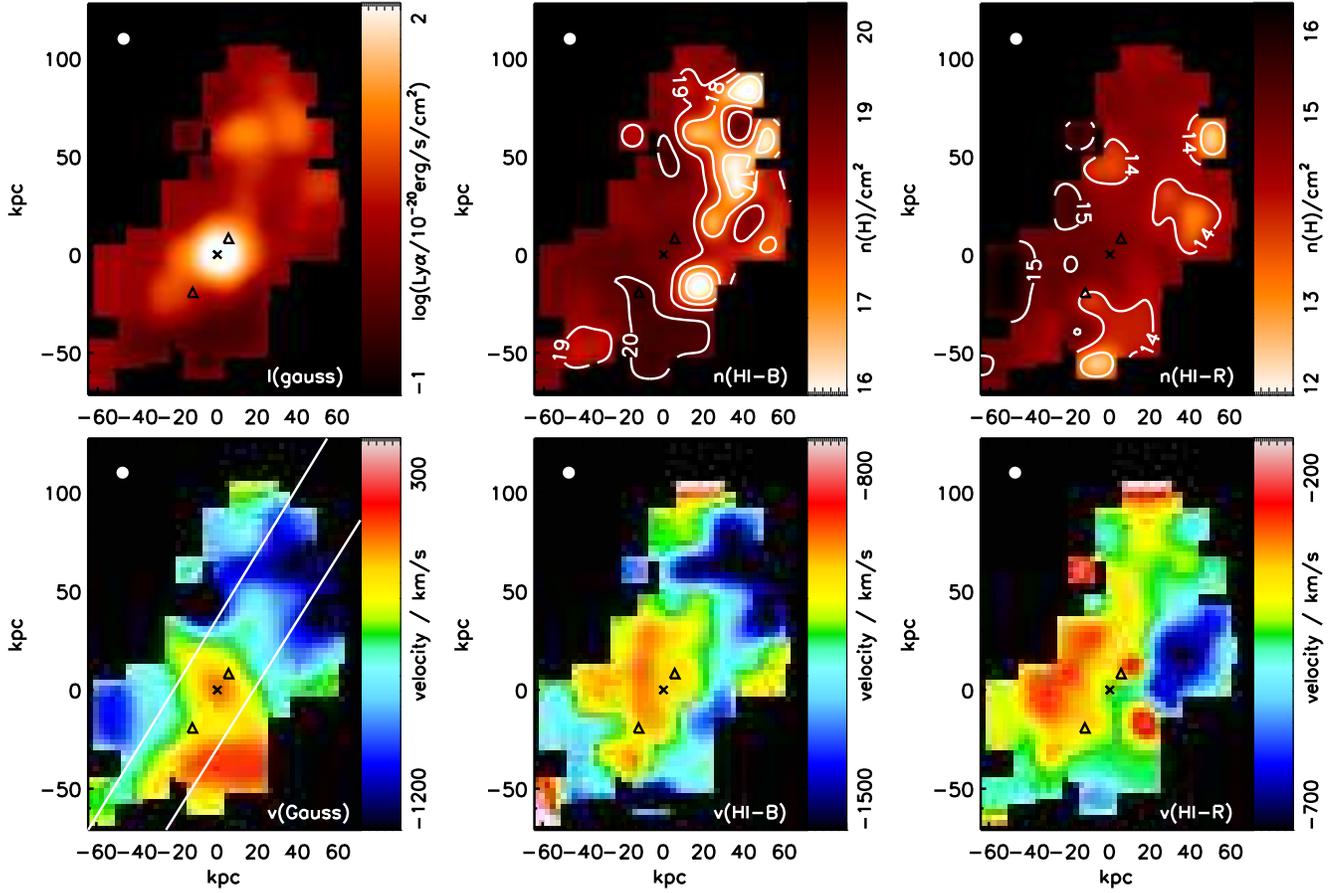,width=7in,angle=0}}
  \vspace{0.3cm}
  \caption{{\it Top row:} Intensity map of the underlying Ly$\alpha$
    Gaussian emission line profile ({\it Left}), and spatial column
    density distribution of H{\sc i}-B ({\it Middle}) and H{\sc i}-R
    ({\it Right}).  {Bottom Row:} Velocity structure of the Gaussian
    profile, H{\sc i}-B and H{\sc i}-R.  In all cases, the velocity
    scale is set by the systemic, as measured from the He{\sc ii}
    emission: $z$\,=\,4.1057.  The solid point in the top left corner
    of each panel denotes the 0.7$''$ FWHM seeing disk.  In each
    panel, we also show the position of the radio core (and continuum
    counterpart of the galaxy; cross) and the radio lobes (open
    triangle).  The underlying Gaussian emission line profile shows a
    maximum velocity gradient of $\sim $\,1000\,km\,s$^{-1}$ across
    the 150\,kpc (in projection).  At the position of the galaxy, the
    Ly$\alpha$ is blue-shifted by $\sim$\,150\,km\,s$^{-1}$ from the
    systemic.  The velocity dispersion profile of the underlying
    emission (which has an average value of
    $\sigma$\,=\,570\,$\pm$\,13\,km\,s$^{-1}$) is peaked at the
    position of the radio galaxy and the radio-jets, which may reflect
    the interaction of the radio plasma and the gas.  Both H{\sc i}-B
    and H{\sc i}-R have covering factors of unity.  H{\sc i}-B appears
    blue-shifted from the (underlying) Gaussian emission line profile
    by $\sim$\,$-$700\,km\,s$^{-1}$ and has an average column density
    of $n$(H{\sc i})$\sim$\,10$^{19.3}$\,cm$^{-2}$.  This material is
    most likely associated with an out-flowing shell of material, and
    we estimate the mass in the shell must be of order
    $\sim$10$^{10}$\,M$_{\odot}$.  The second absorber, H{\sc i}-R is
    much closer to the underlying halo emission in velocity space,
    $\Delta v\sim$\,$-$200\,km\,s$^{-1}$ but has a much lower column
    density, with an average of $n$(H{\sc
      i})$\sim$\,10$^{14.5}$\,cm$^{-2}$.  This material is most likely
    to be associated with the low-density ICM between the out-flowing
    shell and the Ly$\alpha$ halo.}
  \label{fig:nHmaps}
\end{figure*}

\subsection{Spatially Resolved Structure}
\label{sec:resolved}

Before determining the velocity structure of the halo, we first define
the systemic redshift of the galaxy by extracting a spectrum from a
1\,$\times$\,1$''$ region centered on the UV (and radio) continuum
counterpart.  Unfortunately, no strong photospheric absorption
features are present, but there are strong detections of Ly$\alpha$,
C{\sc iv}$\lambda$\,1549 and He{\sc ii}\,$\lambda$\,1640 (see
Fig.~\ref{fig:CoreHaloSpec}).  Since the Ly$\alpha$ and C{\sc iv} are
resonant lines, we use the He{\sc ii} emission line
and fit a single Gaussian profile, deriving a redshift of
$z$\,=\,4.1057\,$\pm$0.0004 and a FWHM of 35\,$\pm$\,3\,\AA\ (or
1250\,$\pm$\,100\,km\,s$^{-1}$).  In all following sections, we adopt
$z$\,=\,4.1057 as the systemic redshift when comparing velocities or
velocity differences.

To investigate the spectral properties of the core and halo, we begin
by extracting two spectra, one from a circular
$\sim$\,3\,$\times$\,3$''$ region centered on the continuum
counterpart (the core of the galaxy), and a second from the extended
halo (excluding the core region), and we show these two spectra in
Fig.~\ref{fig:CoreHaloSpec}.  The core spectrum shows strong emission
lines associated with the AGN (e.g.\ N{\sc v}, [O{\sc iv}], C{\sc iv}
and He{\sc ii}).  The emission from the halo (which should exclude any
seeing-related contamination from the core, but will not exclude
scattered emission within the halo) also shows weak C{\sc iv}
emission.  The detection of heavy elements (in particular C{\sc iv})
in the extended halo suggests that the halo is already metal-enriched
\citep[see also][]{Binette06,VillarMartin03}.

The line ratios from Carbon, Nitrogen or Helium compared to Hydrogen
can be used to infer the physical conditions within the diffuse halo.
Although the C{\sc iv} and He{\sc ii} emission lines lie in
regions of high atmospheric OH emission, we derive line ratios (or
limits) of [C{\sc iv}]\,/\,Ly$\alpha$\,=\,0.082\,$\pm$\,0.003 and
0.064\,$\pm$\,0.013 for the ``core'' and ``halo'' respectively.  For
the He{\sc ii}\,/\,Ly$\alpha$ ratio, we derive
He{\sc
  ii}\,/\,Ly$\alpha$\,=\,0.044\,$\pm$\,0.003 and $<$0.075 for the
``core'' and ``halo'' respectively -- consistent with stellar or
power-law photo-ionisation models with metallicities ranging from
0.05--1\,Z\,/\,Z$_\odot$ \citep{Binette06,VillarMartin07b}.

Next we extract a number of spectra across the halo by integrating
$\sim $\,2$''$\,$\times$\,2$''$ regions and show these in
Fig.~\ref{fig:ImageSpec}.  Whilst the strong Ly$\alpha$ emission is
clearly evident, it is also apparent that the emission line is
asymmetric.  One possible interpretation of the Ly$\alpha$ emission
line profile is that it naturally arises from the transfer of
Ly$\alpha$ radiation through a neutral screen.  If the Ly$\alpha$
photons have to travel through a region of neutral hydrogen, they will
resonantly scatter until they Doppler shift into the wings where the
opacity is lower.  Such a phenomenon will give rise to either a
double-peaked profile \citep{Adams75,Dijkstra06}, or, depending on the
geometry and kinematics of the neutral gas, an asymmetric profile in
which one of the peaks is stronger.  However, for this to be a viable
explanation of the observed line profile in TNJ\,1338, we require the
Ly$\alpha$ to be embedded in a (partial) cocoon of neutral hydrogen of
large optical depth.  On the other hand, the asymmetry of the emission
line, and the sharpness of the deep trough(s) suggest that H{\sc i}
absorption is far more probable
\citep[e.g.\ ][]{Rottgering95,vanojik97,Wilman04}.

Within the central regions of the halo, the asymmetric Ly$\alpha$
profile was first studied using high-resolution UVES spectroscopy by
\citep{Wilman04} who showed that the line emission is best fit with an
underlying Gaussian profile with two H{\sc i} absorbers with column
densities $n$(H{\sc i})\,$\sim$\,10$^{15.0}$\,cm$^{-2}$ and n(H{\sc
  i})\,$\sim$\,10$^{19.7}$\,cm$^{-2}$ for H{\sc i}-B and H{\sc i}-R
respectively.  As we shall see, a similar combination of absorbers is
required to reproduce our spectra across all of the halo.

To describe the line profile, we begin by fitting a Gaussian profile
emission line with a H{\sc i} absorber (which is described by a
Voigt-profile).  However, in all of the regions shown in
Fig.~\ref{fig:ImageSpec}, this fit fails to provide an adequate
description of the line, and instead two H{\sc i} absorbers are
required to adequately describe the line profile \citep[see
  also][]{Wilman04}.  The first of these absorbers, which we label
H{\sc i}-R (the most redshifted) is typically offset from the central
(underlying) Gaussian profile by $\sim $\,$-$270\,km\,s$^{-1}$, has a
column density of $n$(H{\sc i})$\sim$\,10$^{14.5}$\,cm$^{-2}$ and
Doppler parameter of $b$\,=\,100\,$\pm$\,4\,km\,s$^{-1}$ whilst the
second absorber, H{\sc i}-B (the most blue-shifted) is typically
offset from the underlying Gaussian profile by $\sim
$\,$-$1200\,km\,s$^{-1}$ but with a much higher column density,
$n$(H{\sc i})$\sim$\,10$^{19.3}$\,cm$^{-2}$ and
$b$\,=\,170\,$\pm$\,5\,km\,s$^{-1}$.  Within the halo, the underlying
Gaussian profile has an average velocity dispersion of
$\sigma$\,=\,570\,$\pm$\,13\,km\,s$^{-1}$, in agreement with the
velocity dispersions of the He{\sc ii} emission line, and much broader
than the two absorbers.  The velocity dispersion of the underlying
emission is peaked at the position of the radio galaxy (and the
radio-jets), with $\sigma$\,=\,750\,$\pm$\,20\,km\,s$^{-1}$.  This may
reflect two components of the halo: the inner halo where there is a
strong interaction between the radio plasma and the gas, and the outer
(more quiescent) halo, located outside the radio source
\citep{vanojik96}.  In contrast, the much smaller velocity dispersion
of the absorbing gas is more readily explained if these absorbers lie
at relatively large distances from the parent galaxy.  We will discuss
the spatially resolved properties of these absorbers in
\S~\ref{sec:HI-Dynamics}.

Finally, we note that \cite{vanojik97} showed that most high redshift
radio sources with radio jets within $\sim$\,50\,kpc of the central
galaxies show strong H{\sc i} absorption against their Ly$\alpha$
emission.  This may suggest that when the radio sources grow, they
destroy the H{\sc i}.  Since the radio emission in TNJ\,1338 is
located within $\sim$\,10\,kpc of the galaxy, this radio source
clearly falls into first category.  Within this scenario, there must
be a duty cycle with the feedback expelling Ly$\alpha$-emitting gas,
which cools to form a shell of H{\sc i} around it, and a subsequent
feedback event ionises the gas.

\subsection{Dynamics of the H{\sc i} Absorbers}
\label{sec:HI-Dynamics}

As can be seen from Fig.~\ref{fig:ImageSpec}, one of the striking
results from our MUSE observations is the coherence and spatial extent
of both absorbers.  The most blue-shifted absorber, H{\sc i}-B, which
has $n$(H{\sc i})$\sim$\,10$^{19.3}$\,cm$^{-2}$ appears to be
spatially extended across the entire 150\,$\times$\,80\,kpc structure.
The second absorber, H{\sc i}-R, which has a significantly lower
column density, $n$(H{\sc i})$\sim$\,10$^{14.5}$\,cm$^{-2}$ also
appears to be extended over the same area.  Thus, the covering factor
of both absorbers appears to be close to unity.  Evidence for extended
reservoirs of neutral gas 
have been seen in a number of giant Ly$\alpha$ halos around
high-redshift radio galaxies
\citep{vanojik97,Binette00,VillarMartin03,Wilman04} as well as
star-forming galaxies \citep{Bower04,Wilman05,Weijmans10}.

To examine whether the H{\sc i} absorbers are ``pristine'' -- as
suggested by the cold-flow model, in Fig.~\ref{fig:CoreHaloSpec} we
compare the Ly$\alpha$, C{\sc iv} and He{\sc ii} emission line
profiles (transformed to the same rest-frame velocity scale).
Although the C{\sc iv} emission is lower signal-to-noise, it is clear
that the Ly$\alpha$ and C{\sc iv} share the same characteristic
(asymmetric) line profile, implying the presence of C{\sc iv} in the
absorbing material (without other species, it is difficult to derive
the ionisation of the absorbing material, but it appears that this
absorbing gas must be metal enriched, arguing against a model in which
the Ly$\alpha$ results from the cooling of pristine material).

With the two dimensional data provided by MUSE, next, we construct a
spatially resolved map of the dynamics and column density across the
halo.  To achieve this, at each spatial pixel, we average the
surrounding 1$''$\,$\times$\,1$''$ region, and measure the
signal-to-noise (S\,/\,N).  If the S\,/\,N$>$\,50, we then we attempt
to fit the emission line profile with an underlying Gaussian profile
with two H{\sc i} absorbers, allowing the centroid, intensity and
line-width (FWHM in the case of the Gaussian profile and Doppler
$b$-parameter in the Voigt-profile) to be free parameters.  If the
S\,/\,N is not sufficiently high to allow a fit, we adaptively bin up
to 3\,$\times$\,3$''$ (in steps of 0.2$''$) and re-fit.

\begin{figure}
  \centerline{\psfig{file=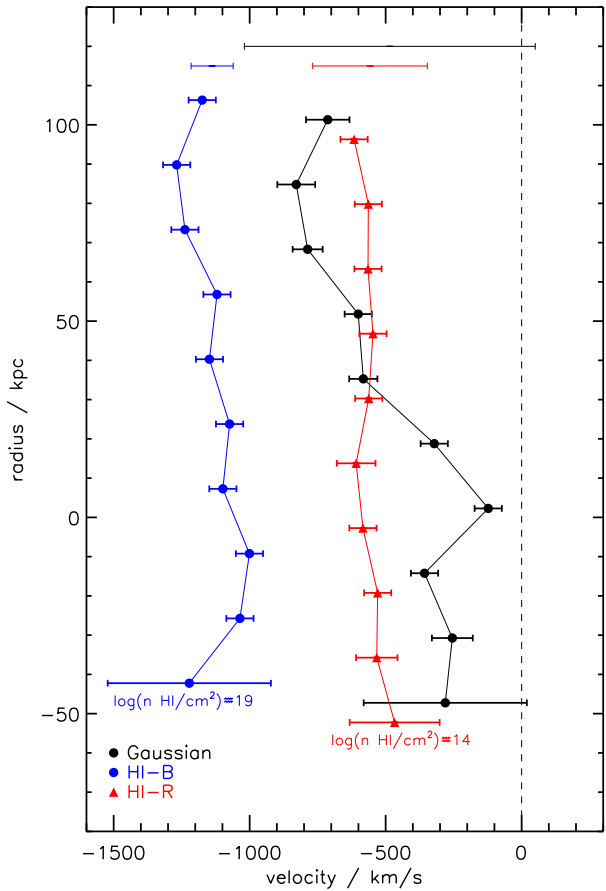,width=3.5in,angle=0}}
  \caption{The velocity profile of the underlying Gaussian emission
    and the two absorbers extracted across the major axis of the
    Ly$\alpha$ halo (as shown in Fig.~\ref{fig:nHmaps}). The solid
    points and thick error bars denote the central velocity and the
    error on the central velocity, whilst the thinner errors at the
    top of the panel show the average width of the Gaussian ($\sigma$)
    or Doppler $b$-parameter (absorbers) in each bin.  The origin of
    the velocity and spatial axes is fixed on the radio galaxy,
    demonstrating that these motions do not reflect bulk {\it
      rotation} of gas around the radio galaxy if it lies at the
    bottom of the potential well \citep[c.f.\ ][]{vanojik97}.}
    \label{fig:1dRC}
\end{figure}

In Fig.~\ref{fig:ImageSpec} we show the velocity field of the
underlying Gaussian emission, as well as that of the two H{\sc i}
absorbers.  The underlying emission has a strong velocity gradient
($\sim $\,700\,km\,s$^{-1}$) across the 150\,kpc in projection.  At
the position of the galaxy, the Ly$\alpha$ is blue-shifted by
$\sim$\,150\,km\,s$^{-1}$ from the systemic, and this velocity offset
decreases up to $\sim$\,1200\,km\,s$^{-1}$ towards to North-West.  The
H{\sc i} absorber which has a velocity closest to the underlying
emission, H{\sc i}-R, does not have a strong velocity gradient across
the major axis.  However, the most blue shifted absorber, H{\sc i}-B
is blue-shifted from the underlying emission by
$\sim$\,700\,km\,s$^{-1}$ with a velocity field that apparently
mirrors that of the underlying Gaussian profile.  To highlight these
differences, we extract the dynamics through a pseudo-slit from the
major kinematic axis of the halo and show this in Fig.~\ref{fig:1dRC}.

Turning to the column density maps, Fig.~\ref{fig:nHmaps} shows that
both of the H{\sc i} columns are remarkably uniform (although at two
distinct velocities), with both remaining as single systems of the
full extent of the Ly$\alpha$ emission.  The spatial and kinematic
coherence of the absorbers suggest they may be physically distinct
from the background source.  Using the column densities and sizes, the
total {\it neutral} mass of H{\sc i}-B and H{\sc i}-R are M$_{\rm
  HI-B}\sim $\,10$^9$\,M$_{\odot}$ and M$_{\rm HI-R}\sim
$\,10$^5$\,M$_{\odot}$.  The {\it total} hydrogen mass of the absorber
depends on the ionisation state.  Photoionisation models of absorption
systems with column densities in the range $n$(H{\sc
  i})\,=\,10$^{19-20}$\,cm$^{-2}$ suggest average IGM neutral
fractions at $z$\,=\,4 of $X_{\rm HI}\sim\,$1.5--10 \citep{McQuinn11}.
Of course, these values may represent upper limits in an over-dense
region which self-shields from UV radiation.  Nevertheless, this
suggests a total mass for the most massive absorber, H{\sc i}-B of
$\sim$10$^{10}$\,M$_{\odot}$.

\subsubsection{Inflow or Outflow?}

Previous observations of high-redshift radio galaxies have identified
very strong (high column density) absorbers seen against the
Ly$\alpha$ profiles \citep[][]{Hippelein93,vanojik97,Humphrey07}.  The
interpretation of this absorption is the fortuitous alignment with the
intervening Ly$\alpha$ forest absorption, which may be enhanced in the
biased regions around massive, high-redshift galaxies.  Whilst the
properties of absorbing halos around radio galaxies have been studied
in detail, most studies have been limited to (either) low resolution
spectra and\,/\,or integrated spectra, but where it has been possible
to spatially resolve the absorption, the implied covering factor is
close to unity -- on scales from 10--50\,kpc -- and the inferred mass
of the neutral phase is $\sim$\,10$^7$--10$^8$\,M$_{\odot}$.  The
absorbers also appear to be ubiquitous when seen against radio quiet
galaxies, such as the giant ``Ly$\alpha$ blobs'' (LABs) at $z$\,=\,3.1
which have also been studied in detail \citep{Matsuda11} (we note that
the other absorber in this field also shows a strong, spatially
extended absorber in the MUSE data; Urrutia et al.\ in prep).  For
example, \citet{Matsuda06} find that the fraction of LABs with
absorbers seen against the Ly$\alpha$ emission is 83\%.  However, the
nature of the absorbing gas remains poorly understood even when
two-dimensional spectra are available.  In particular, two-dimensional
spectra are required to test whether the neutral gas is likely to be
in-falling or out-flowing.  With sufficient signal to noise and spectral
(and spatial) resolution the spatially resolved dynamics of the
neutral gas provided by MUSE, and its relation to the (underlying)
emission should allow us to distinguish between infall and outflow,
and we consider both possibilities here.

In a scenario in which the absorber(s) around TNJ\,1338 arise due to
foreground material, the giant Ly$\alpha$ halo only provides the
background (illuminating) screen from which the absorbers are
decoupled.  In this case, the coherence of absorbers may suggest we
are seeing a sheet (or filament) with a projected separations from the
radio galaxy of $\sim$\,4\,Mpc and $<$\,1Mpc (assuming a Hubble
constant at $z$\,=\,4.1 of H($z$)\,=\,423\,km\,s$^{-1}$\,Mpc).  The low
column density absorber, H{\sc i}-R is presumably within the virial
radius of the halo associated with the radio galaxy since the velocity
offset places it within $\sim$\,1\,Mpc of the galaxy.

Whilst this interpretation is appealing since high density IGM is
expected to be seen around massive galaxies, there are a number of
problems with this inflow model.  First, the H{\sc i} appear to be
enriched, although this does not preclude them being unassociated with
the radio galaxy, especially in an over-dense (biased) region.
Second, the velocity structure of the underlying Ly$\alpha$ emission
(which has an average velocity gradient across the halo of
dv\,/\,dR$\sim $\,4\,km\,s$^{-1}$\,kpc$^{-1}$) is mirrored by H{\sc
  i}-B (dv\,/\,dR$\sim $\,3\,km\,s$^{-1}$\,kpc$^{-1}$) strongly
suggesting that the material associated with H{\sc i}-B originated in
the Ly$\alpha$ halo (which would then naturally explain the similarity
in the Ly$\alpha$ and C{\sc iv} line profiles).  Third, across the
halo, the low density absorber, H{\sc i}-R, has a median Doppler
$b$-parameter of $b$\,=\,170\,$\pm$\,3\,km\,s$^{-1}$.  We can test if
this is ``typical'' of Ly$\alpha$ absorbers by using quasar sight
lines.  \citet{Fechner07} and \citet{Schaye00} derive column density
and Doppler $b$ parameters for neutral hydrogen in the Ly$\alpha$
forest at $z$\,=\,2--4.5.  In total, they detect $\sim$\,1500
absorbers with columns between log\,$n$(H{\sc
  i}\,/\,cm$^{-2}$)\,=\,12--17 and $b$\,=\,5--200\,km\,s$^{-2}$.
Cutting their sample to the same median column density as seen in
H{\sc i}-R, log\,$n$(H{\sc i}\,/\,cm$^{-2}$)\,=\,14.5, the median
Doppler parameter is $b$\,=\,30\,$\pm$\,1\,km\,s$^{-1}$, a factor
5.5\,$\pm$\,0.4 lower (for the same median column density) as that
seen in H{\sc i}-R.  Thus, it appears that the low column density
absorber seen against the halo of TNJ\,1338 does not have the
properties expected for simply being ``average'' IGM, but instead has
a much higher velocity width.

A more natural explanation for the absorbers is therefore that H{\sc
  i}-B represents the out-flowing material swept up in a wind shell
ejected from the host galaxy and H{\sc i}-R is the low density
material between this shell and the host.  Galactic scale outflows
from high-redshift galaxies, driven by the collective effects of
star-formation and supernovae winds
\citep[e.g.\ ][]{Pettini02a,Bower04,Erb06c,Swinbank07a,Swinbank09,Steidel10,Martin12,Newman12}
and\,/\,or AGN
\citep[e.g.\ ][]{Nesvadba07,Alexander10,Martin12,Harrison12,Harrison14,Genzel14}
have been studied for some time and outflow velocities reaching
$\sim$1000\,km\,s$^{-1}$ in the ionised gas spread over tens of
kilo-parsecs.

Can we explain the properties of the absorbers in such a model?
Numerical models can be used to trace the evolution of the ISM
following the energy injection into an ambient medium, thus making
predictions for the effect of a starburst or AGN-driven wind on the
ISM.  For example, \citet{Krause05} use a hydro-dynamic code to model
a galactic wind (due to star formation and supernova) on the ISM of a
host galaxy and superimpose an AGN-driven jet.  Assuming the wind
starts long before the jet activity, the starburst first drives the
wind in to a radiative bow shock, which then cools and is seen in
Ly$\alpha$ absorption.  When the jet reaches the shell, it may
destroys it (at least partially) so that the larger sources are no
longer absorbed (the shell eventually cools and fragments to form
globular clusters).  In this model, the power required to drive the
wind is approximately
L\,$\gsim$\,5\,$\times$\,10$^{43}$\,erg\,/\,s\,$\sqrt{T/10^6 K}$
($v_{\rm shell}$\,/\,200\,km\,s$^{-1}$)$^{4}$\,($r_{\rm
  shell}$\,/\,25\,kpc).  For the outflow in TNJ\,1338, with $v_{\rm
  shell}\sim$\,1000\,km\,s$^{-1}$ and $r_{\rm shell}\sim$\,200\,kpc,
L\,$\sim$\,1\,$\times$\,10$^{46}$\,erg\,s$^{-1}$.  Although this
calculation should be considered approximate, the luminosity required
to drive the shell is consistent with the AGN luminosity of TNJ\,1338,
L$_{\rm AGN}\sim$\,2\,$\times$\,10$^{46}$\,erg\,s$^{-1}$
\citep{Drouart14}.

Another approach is to consider the energy of the AGN (or star
formation) and the likely time-scales involved.  Assuming the outflow
in TNJ\,1338 is a coherent structure, the mass swept up in the shell
must be of order $M_{\rm outflow}\sim$\,10$^{10}$\,M$_{\odot}$ (for a
neutral fraction of 10\%) and it will have a kinetic energy of
$\sim$10$^{59}$\,erg (assuming a spherical shell with mass
$\sim$\,10$^{10}$\,M$_{\odot}$ moving at 1200\,km\,s$^{-1}$ with
respect to the galaxy).  If the shell is located a minimum of
$\sim$\,200\,kpc from the radio galaxy (this minimum being set by the
spatial extent of the Ly$\alpha$ halo and coherence of the absorber),
then the {\it minimum} lifetime of the outflow is 100\,Myr.
Considering the AGN luminosity of this source, L$_{\rm
  AGN}\sim$\,2\,$\times$\,10$^{46}$\,erg\,s$^{-1}$ and a minimum
timescale of 100\,Myr, the total energy available is
$\sim$\,5\,$\times$\,10$^{61}$\,erg.  Alternatively, considering star
formation as the energy injection mechanism and assuming supernovae
provide 10$^{49}$\,erg per solar mass of stars formed, the energy
provided by SNe will be $\sim$\,10$^{60}$\,erg.  Thus energetically, this
star-formation or AGN could drive the wind if they are able to couple
$<$\,10\% of their energy to the wind (and the energy in the wind
should be considered a lower limit since the galaxy--shell distance is
a lower limit).

We can also compare the mass in the absorber to the
  total ISM mass of the host galaxy.  Fitting a modified black-body to
  the {\it Herschel} 250--500$\mu$m, SCUBA\,850$\mu$m and
  MAMBO\,1.2\,mm photometry of this galaxy
  \citep{DeBreuck04,Drouart14}, adopting a dust emissivity of of
  $\beta$\,=\,1.5--2.0 \citep{Magnelli12} and dust mass absorption
  coefficient of $\kappa_{\rm 870\mu m}$\,=\,0.15\,m$^2$\,kg$^{-1}$
  \citep{WeingartnerDraine01}, we estimate a cold dust mass of M$_{\rm
    dust}\sim $\,3\,$\times$\,10$^8$\,M$_{\odot}$. For a dust-to-gas
  ratio of $\delta_{\rm DGR}$\,=\,90 \citep{Bothwell13,Swinbank14},
  this suggests a H$_{\rm 2}$ mass of M$_{\rm
    H_2}\sim$\,3\,$\times$\,10$^{10}$\,M$_{\odot}$, which is
  comparable to that estimated for other HzRGs
  \citep{Emonts14,Miley08}.  We caution that there is a factor
  3--4\,$\times$ uncertainty on this value due to the (moderately low)
  signal-to-noise of the 850\,$\mu$m and 1.2\,mm detections (and upper
  limits on the 250, 350 and 500\,$\mu$m photometry), the choice of
  dust model parameters, and the uncertainty in the adopted
  dust-to-gas ratio.  Nevertheless, the H$_2$ mass of the ISM in the
  radio galaxy is similar to that estimated for the total mass of the
  two structures we see in absorption against the halo.

Will the gas escape the halo potential?  If we consider a
$\sim$10$^{11}$\,M$_{\odot}$ galaxy inside a dark halo with mass
$\sim$10$^{13}$--10$^{14}$ with NFW \citep{NFW} density profile, the
gas would require a velocity greater than $\sim$\,1000\,km\,s$^{-1}$
to escape the halo (see also the calculation in \citealt{Harrison12}).
This is similar to the velocity seen in H{\sc i}-B.  However, some
models have suggested that even massive outflows may stall in the
galaxy halo (or in this case, the ``group'' halo), and re-collapse and
cool at later times (along with new fuel supplies), resulting in
re-ignition of star formation and further black-hole growth
\citep[e.g.\ ][]{Lagos08,Gabor11,Hopkins13}, especially in
group/cluster (or proto-cluster) environments \citep{McCarthy11}.

In summary, while it is not possible to constrain definitively whether
the AGN or star-formation have driven the outflow, or the ultimate
fate of out-flowing gas, it appears that the H{\sc i} absorbers are
associated with activity occurring within the galaxy, and with outflow
properties (high velocity, spatially extended, and mass loaded) in
broad agreement with the requirements of models that invoke AGN-driven
outflows to regulate star formation and black-hole growth in luminous
AGN \citep[][]{DiMatteo05,Hopkins10,Debuhr12}.  Even if the outflow
does not escape the galaxy halos, it may have sufficient energy to
heat the halo and control the level of cooling in massive halos at
later times \citep[e.g.\ ][]{Churazov05,Gabor11,McCarthy11,Bower12}.

%
%

\section{Conclusions}

We have used the MUSE integral field spectrograph to study the giant
Ly$\alpha$ has around the high-redshift radio galaxy TNJ\,1338 at
$z$\,=\,4.11.  Our main observations reveal the following:
\smallskip

\noindent $\bullet$ Our observations map the intensity distribution and dynamics
of the Ly$\alpha$ halo which is extended over a region approximately
150\,$\times$\,80\,kpc.  In comparison, the radio jets are confined to
the central $\sim$\,10\,kpc of the halo, although the alignment of the
extended Ly$\alpha$ halo and radio jets may be indicative of previous
AGN activity (which ionised the extended halo and which is now seen in
Ly$\alpha$ emission as the gas cools).
\smallskip

\noindent $\bullet$ The spatial extent of the Ly$\alpha$ halo in TNJ\,J1338 is
one of the largest seen in high-redshift radio galaxies.  We clearly
detect the C{\sc iv} emission in the extended halo suggesting that the
gas in the extended is enriched.
\smallskip

\noindent$ \bullet$ Across the entire halo, the Ly$\alpha$ emission
line appears asymmetric.  The line profile is best-fit by two H{\sc-i}
absorbers superimposed on an underlying Gaussian emission profile.
One of the striking results from our MUSE observations is the
coherence and spatial extent of both absorbers.  Indeed, both
absorbers have a covering factor of unity.  The first absorber, H{\sc
  i}-B appears blue-shifted from the (underlying) Gaussian emission
line profile by $\sim$\,$-$700\,km\,s$^{-1}$ and has a column density
of $n$(H{\sc i})$\sim$\,10$^{19.3}$\,cm$^{-2}$.  Since the velocity
structure of this absorber follows that of the underlying emission,
this material is most likely associated with an out-flowing shell of
material, and we estimate the mass in the shell must be of order
$\sim$10$^{10}$\,M$_{\odot}$.  By comparison to the AGN luminosity or
star-formation rate of the galaxy, we find that both could trivially
drive the wind if they are able to couple $<$\,10\% of their energy to
the wind.  The second absorber, H{\sc i}-R is much closer to the
underlying halo emission, $\Delta v\sim$\,$-$200\,km\,s$^{-1}$ but has
a much lower column density, $n$(H{\sc
  i})$\sim$\,10$^{14.5}$\,cm$^{-2}$.  The Doppler $b$ parameter for
this absorber is $\sim$6$\times$ higher than expected for the
low-density IGM (at the same column density), suggesting that this is
not just a typical intergalactic absorber, but instead that this
material represents the low-density IGM between the out-flowing shell
and the Ly$\alpha$ halo.  We show that the energetics of the
star-formation or AGN could drive the outflow if it is able to couple
a small fraction of the energy to the gas.
\smallskip

\noindent$ \bullet$ The underlying Gaussian emission line profile shows a
velocity gradient of $\sim $\,700\,km\,s$^{-1}$ across the 150\,kpc in
projection.  At the position of the galaxy, the Ly$\alpha$ is
blue-shifted by $\sim$\,150\,km\,s$^{-1}$ from the systemic.  The
velocity dispersion profile of the underlying emission (which has an
average value of $\sigma$\,=\,570\,$\pm$\,13\,km\,s$^{-1}$) is peaked
at the position of the radio galaxy and the radio-jets, which may
reflect the interaction of the radio plasma and the gas.
\smallskip

\noindent$ \bullet$ We identify two other $z$\,=\,4.1 emitters in the field of
the radio galaxy which are presumably associated with the
group/cluster within this over-density.  We also serendipitously
identify a new giant Ly$\alpha$ halo at $z$\,=\,3.31 approximately
$\sim$\,30$''$ to the south west of TNJ\,J1338 which appears to be
associated with a number of other star-forming galaxies and AGN at
this redshift.
\smallskip

Finally, we can consider how the galaxy and outflow may evolve towards
the present day.  The out-flowing shell will be both thermally and
gravitationally unstable, and will eventually fragment.  At this
point, the covering factor will decrease and the absorption will
vanish (which would explain why no absorbers are seen in the halos of
radio galaxies with jet activity $>$\,50\,kpc scales;
\citealt{Rottgering99conf}).  The Jeans mass in the shell will be of
order 10$^6$\,M$_{\odot}$ \citep{Krause02}.  Hence the shell will
fragment and form stars in globular clusters of this mass.  Of course,
at this point the shell should become visible in the rest-frame
UV\,/\,optical as young stars form (which will also increase the
ionisation of the remaining parts of the shell).  Assuming an
efficiency of $\sim$\,10\%, this shell will contain a $\sim$\,1000
globular clusters of this mass -- in reasonable agreement with the
number of clusters found in nearby brightest cluster galaxies
\citep[][]{Harris98} which show excess of globular clusters systems
compared to non-BCGs of comparable luminosity.

In summary, the most successful models of galaxy formation
increasingly require outflows of gas to expel gas from galaxies at
early times.  In the most massive galaxies, most of the energy
injection is believed to arise from the central AGN, making
high-redshift radio galaxies, which have many of the characteristics
expected for the progenitors of todays massive Ellipticals, strong
candidates to empirically constrain the properties of outflows which
can be used to test the models.  The discovery of high-velocity
($\sim$\,1000\,km\,s$^{-1}$), halo-scale (i.e. $\gg$\,150kpc) and mass
loaded winds in the vicinity of the central radio source are broadly
in agreement with the requirements of models that invoke AGN-driven
outflows to regulate star formation and black-hole growth in luminous
AGN \citep[][]{DiMatteo05,Hopkins10,Debuhr12}.

\section*{Acknowledgments}

We thank the referee for the constructive report on this paper and
Yuichi Matsuda, Michele Fumagalli and Tom Theuns for useful
discussions.  We would also like to thank the pipeline developers for
the data reduction package.  AMS gratefully acknowledges an STFC
Advanced Fellowship through grant number ST/H005234/1 and the
Leverhume foundation.  IRS acknowledges support from STFC
(ST/I001573/1), the ERC Advanced Investigator programme DUSTYGAL
321334 and a Royal Society/Wolfson Merit Award.  RB acknowledges
support from the ERC advanced grant 339659-MUSICOS.  BPV acknowledges
funding through ERC grant ``Cosmic Dawn''.  This publication uses data
taken from the MUSE science verification programme 60.A-9318 and
commissioning run 060.A-9100.  All of the data used in this paper is
available through the ESO science archive.


\end{document}